Original Paper

# Promoting Health via mHealth Applications Using a French Version of the Mobile App Rating Scale: Adaptation and Validation Study

Ina Saliasi[1*], PhD; Prescilla Martinon[1*], MSc; Emily Darlington[1], PhD; Colette Smentek[1], PhD; Delphine Tardivo[2], PhD; Denis Bourgeois[1], PhD; Claude Dussart[1], PhD; Florence Carrouel[1*], PhD; Laurie Fraticelli[1*], PhD

[1]Health, Systemic, Process, Research Unit 4129, University Claude Bernard Lyon 1, University of Lyon, Lyon, France
[2]Laboratory Anthropology, Health Law, and Medical Ethics, UMR 7268, Aix-Marseille University 2, Marseille, France
[*]these authors contributed equally

**Corresponding Author:**
Ina Saliasi, PhD
Health, Systemic, Process, Research Unit 4129
University Claude Bernard Lyon 1
University of Lyon
11 rue Guillaume Paradin
Lyon, Cedex 08
France
Phone: 33 0478785745
Email: ina.saliasi@univ-lyon.fr

## Abstract

**Background:** In the recent decades, the number of apps promoting health behaviors and health-related strategies and interventions has increased alongside the number of smartphone users. Nevertheless, the validity process for measuring and reporting app quality remains unsatisfactory for health professionals and end users and represents a public health concern. The Mobile Application Rating Scale (MARS) is a tool validated and widely used in the scientific literature to evaluate and compare mHealth app functionalities. However, MARS is not adapted to the French culture nor to the language.

**Objective:** This study aims to translate, adapt, and validate the equivalent French version of MARS (ie, MARS-F).

**Methods:** The original MARS was first translated to French by two independent bilingual scientists, and their common version was blind back-translated twice by two native English speakers, culminating in a final well-established MARS-F. Its comprehensibility was then evaluated by 6 individuals (3 researchers and 3 nonacademics), and the final MARS-F version was created. Two bilingual raters independently completed the evaluation of 63 apps using MARS and MARS-F. Interrater reliability was assessed using intraclass correlation coefficients. In addition, internal consistency and validity of both scales were assessed. Mokken scale analysis was used to investigate the scalability of both MARS and MARS-F.

**Results:** MARS-F had a good alignment with the original MARS, with properties comparable between the two scales. The correlation coefficients ($r$) between the corresponding dimensions of MARS and MARS-F ranged from 0.97 to 0.99. The internal consistencies of the MARS-F dimensions *engagement* (ω=0.79), *functionality* (ω=0.79), *esthetics* (ω=0.78), and *information quality* (ω=0.61) were acceptable and that for the overall MARS score (ω=0.86) was good. Mokken scale analysis revealed a strong scalability for MARS (Loevinger H=0.37) and a good scalability for MARS-F (H=0.35).

**Conclusions:** MARS-F is a valid tool, and it would serve as a crucial aid for researchers, health care professionals, public health authorities, and interested third parties, to assess the quality of mHealth apps in French-speaking countries.

## Introduction

In the last few decades, smartphones have radically modified our daily life, as seen by the increasing number of smartphone users worldwide. In parallel to this, an exponential growth of mobile health (mHealth) apps has been observed [1]. Such apps offer an attractive and promising interface for health education and community health promotion [2]. mHealth apps are currently becoming handheld devices that can disseminate a variety of health-promoting knowledge and promote healthy behaviors relating, for example, to dietary habits [3], weight control [4], physical activity [5], addictive behaviors (ie, smoking), and mental health (ie, managing stress and depression) [1]. mHealth apps represent an alternative to or complement face-to-face communication between health care professionals and users of the health care system for primary prevention [6], as well as patients for secondary prevention [7]. They offer an affordable platform that reaches a large audience with possible positive implications for public health, especially health promotion and prevention strategies [1].

Before the deployment of an app on the web, the app store reviews it as well as its updates, in order to determine whether it is reliable, performs as expected, respects user privacy, and is free of objectionable content such as offensive language or nudity. However, the review by the developer is not comprehensive enough to enable end users, health professionals, and researchers to identify and evaluate the quality of mHealth apps [8,9]. The most common way to select an mHealth app that is currently available on the app market is by using publicly available information, and by considering easily available attributes such as title, price, star ratings, reviews, or downloads, instead of validated scientific content [10]. To date, certification and trust labels for mobile apps are not widely endorsed [11].

Few mHealth apps available on the market have undergone a thorough validation process based on high-level evidence that can be a potential problem for the safety of end users [9]. In order to evaluate the validity and functionality of mHealth apps objectively, several standardized scales have been developed for health care professionals [12]. The Mobile Application Rating Scale (MARS) was developed by Stoyanov et al [8] in the English language, and, to date, it is considered the reference scale for health care professionals in the scientific literature. The Italian, Spanish, German, and Arabic versions of MARS have already been produced and validated [2,13-15]. The 23-item scale assesses the quality of health-related apps through four objective dimensions relating to the quality of the mHealth app (engagement, functionality, esthetics, and information) and one subjective dimension (subjective app quality and perceived impact).

The aim of this study is to develop and validate a French version of the Mobile App Rating Scale (MARS-F) as a multidimensional measure for trialing, classifying, and rating the quality of mHealth apps.

## Methods

### Study Design

The validation of this study followed and applied a well-established process of cross-cultural adaptation [16], translation and back-translation, review, piloting, and psychometric evaluation.

### Cultural Adaptation and Translation

First, the translation of MARS from English to French was conducted by two independent bilingual scientists (IS and LF). Following the review, discussion, and comparison of their two forward translations, they agreed upon a common pilot version of MARS-F. Second, this common pilot version was blind back-translated by two bilingual native English speakers with different educational backgrounds—a researcher in public health and educational sciences (ED) and a nonacademic professional (ADB). Third, the two bilingual scientists (IS and LF) compared the back-translated version with the original English version. After mutual discussion, they agreed upon the final French version of the scale (MARS-F). Finally, 6 other people (3 researchers and 3 nonacademic professionals) evaluated the comprehensibility of this finalized French version. Their comments were considered, and the final MARS-F version was thus created (Multimedia Appendix 1).

### Selection of Apps

The inclusion process consisted of three different phases: searching, screening, and determining the eligibility criteria of nutrition health-related apps. The search for apps was conducted from March 10, 2021, to March 17, 2021, on the French Apple Store (iOS) and Google Play Store (Android). No truncation or use of logic operators (AND, OR, and NOT) was possible while searching in the Google Play Store and iOS Store. Hence, in order to select the nutrition health-related apps, the following search terms were used separately: "nutrition" (nutrition), "diététique" (dietetics), "alimentation" (food intake), "régime alimentaire" (diet), and "manger sain" (healthy eating). Apps were included if they were available free of charge or at least free of charge during 7 days from both the iOS Store and Google Play Store. Duplicate copies of apps between the two stores were excluded, resulting in a total of 63 apps (Figure 1).

**Figure 1.** Flowchart of the app selection process.

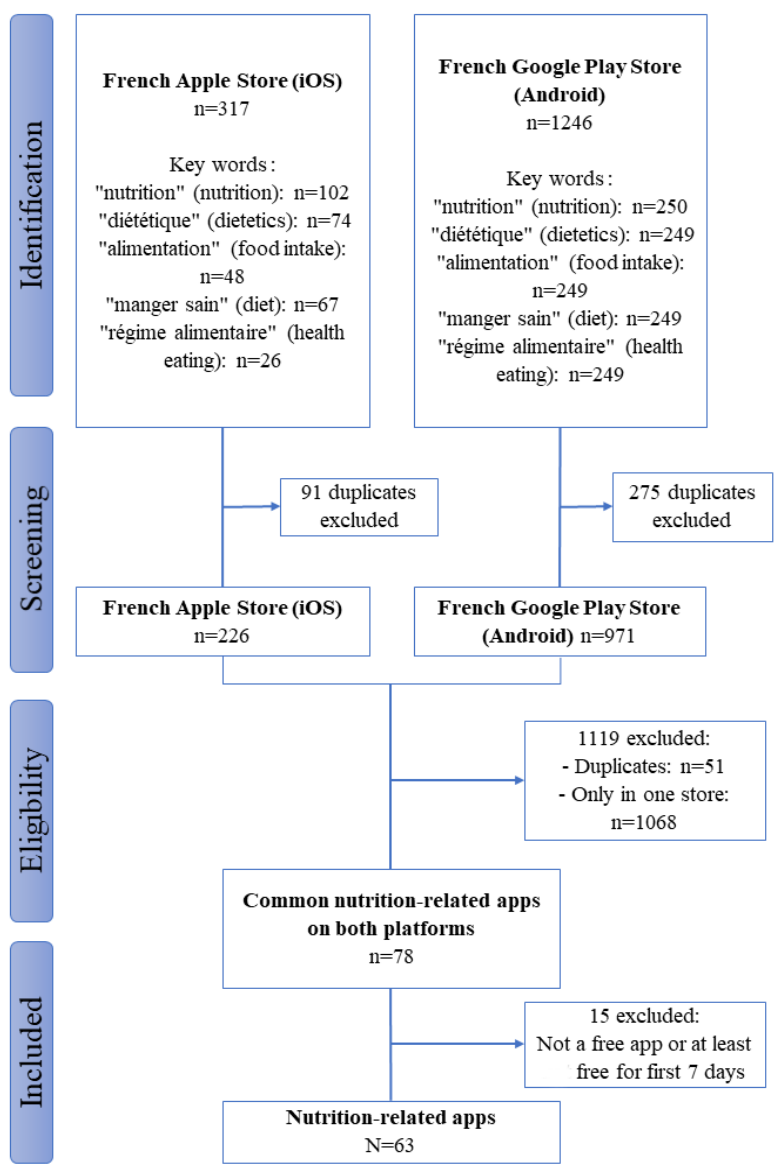

## Raters' Training

To complete the evaluation process of apps, we used the rating methodology previously described by Stoyanov et al [8]. We made a video with an introduction of the French MARS scale, and an exercise on how to rate a nutrition mobile app (available on request to the corresponding author). Two individuals with a master's degree in medical sciences (FC and PM), and who were fluent in both French and English, were instructed on how to serve as raters by watching the video. With a view to ensure that the raters were sufficiently trained, they were asked to download and evaluate 10 apps that were randomly selected from those meeting our inclusion criteria using MARS and MARS-F. Each app rater tested each app for at least 15 minutes before they carried out their evaluation. Raters then compared their individual rating scores for each app. When their individual rating scores varied by at least 2 points, they discussed their findings until they aligned their rating approaches and agreed on the score.

## Data Analysis

### Intraclass Correlation

The two raters completed the evaluation of the remaining 53 apps independently. The intraclass correlation coefficients (ICCs) were calculated to measure the interrater reliability of the items, the subscales, and total MARS scores with absolute agreement between the raters. An ICC of <0.50 was interpreted as poor; 0.51-0.75, as moderate; 0.76-0.89, as good; and >0.90, as excellent correlation [17]. We excluded item 19 due to missing values.

### Internal Consistency

The internal consistency of MARS-F and its subscales were also assessed as a measure of scale reliability, as reported in the original MARS study. We used the omega coefficient instead of the Cronbach alpha coefficient, as it is commonly used to assess reliability as described in the literature. The omega coefficient provides justifiably higher estimates of reliability than the Cronbach alpha coefficient [18]. The robust procedure introduced by Zhang and Yuan was used to estimate omega values, with the objective to obtain a closer estimate to the population value without being overwhelmingly affected by a few overbearing observations [19]. Reliability was assessed as follows: ω<0.50 was interpreted as unacceptable internal consistency; ω=0.51-0.59, as poor consistency; ω=0.60-0.69, as questionable consistency; ω=0.70-0.79, as acceptable consistency; ω=0.80-0.89, as good consistency; and ω>0.90, as excellent consistency.

### Validity

To establish an indicator of validity, we investigated the subscale correlations between MARS-F and its original English version. In addition, we calculated the overall correlation between the total MARS score and total MARS-F score. The correlation coefficient ranges between –1 and 1. The closer the coefficient is to 1, the stronger the positive linear relationship between the variables. The closer the coefficient is to –1, the stronger the negative linear relationship between the variables. Mean comparisons were also performed between the corresponding dimensions of MARS and MARS-F, and *P* values were adjusted for multiple testing according to Holmes' method [20]. For all dimensions compared, we considered a *P* value <.05 as statistically significant.

### Mokken Scale Analysis

Mokken scale analysis (MSA) is a technique used for scaling test and questionnaire data closely. This technique is related to the nonparametric item response theory [21,22]. The latent monotonicity and nonintersection are two necessary preconditions to use the MSA. Loevinger H is the key parameter of this scale. For item $i$, the scaling parameter is $H_i$, and $H_k$ is the scaling parameter for the overall scalability of all items in the scale k. $H_i$ indicates the strength of the relationship between a latent variable (eg, app quality) and item $i$. The SE values of the scalability coefficients of the item pairs were also calculated. A scale is considered weak if H is <0.4, moderate if H is ≥0.4 but <0.49, and strong if H is >0.5 [21,23]. MSA was conducted for both MARS and MARS-F to assess the scalability of the mean scores. As recommended by van der Ark, the reliability of the scales was additionally assessed using the Molenaar-Sijtsma (MS) method [24], λ-2, and latent class reliability coefficient (LCRC) [14,25].

### Statistical Analysis

R software (version 4.0.5; R Foundation for Statistical Computing) was used for all analyses. The correlations, ICC, and MSA were conducted using the R packages psych (function corr.test) (version 1.8.12), coefficient alpha (function omega) (version 0.5) and mokken (function coefH) (version 1.8.12). The two preconditions of latent monotonicity and nonintersection were tested using the functions check.monotonicity and check.restscore from the package mokken. The statistics related to the reliability of the scales were provided using the function check.reliability.

## Results

### Calibration During Raters' Training

Among the 10 common apps, the mean scores of the dimensions *engagement* ($t_{10}$=–0.76, *P*=.44), *functionality* ($t_{53}$=–0.11, *P*=.90), *esthetics* ($t_{53}$=0.22, *P*=.82), and *information quality* ($t_{33}$=–0.35, *P*=.72) were equivalent in both versions. The internal consistencies of MARS (ω=0.86, 95% CI 0.56-0.96) and MARS-F (ω=0.78, 95% CI 0.32-0.94) were good and acceptable, respectively, and the MSA revealed strong scalability (H=0.47; SE=0.07).

### Descriptive Data and Mean Comparisons

The ICCs for MARS (0.88, 95% CI 0.79-0.93) and MARS-F (0.89, 95% CI 0.8-0.93) were high. The mean and SD scores of the items in MARS and MARS-F are presented in Table 1. The mean scores of the dimensions *engagement* ($t_{53}$=–0.34, *P*=.72), *functionality* ($t_{53}$=–0.47, *P*=.63), *esthetics* ($t_{53}$=.09, *P*=.92), and *information quality* ($t_{33}$=0, *P*>.99) were equivalent between MARS and MARS-F.

Table 1. Summary of item and scale scores for the original version of the Mobile App Rating Scale (MARS) and the French version of MARS (MARS-F) (n=53 apps).

| Dimension | Score, mean (SD) | |
|---|---|---|
| | MARS | MARS-F |
| **Engagement** | 2.83 (0.89) | 2.85 (0.88) |
| Item 1 | 2.67 (0.81) | 2.63 (0.82) |
| Item 2 | 2.97 (0.74) | 2.90 (0.78) |
| Item 3 | 2.65 (0.94) | 2.66 (0.92) |
| Item 4 | 2.60 (0.99) | 2.84 (0.95) |
| Item 5 | 3.24 (0.80) | 3.20 (0.79) |
| **Functionality** | 4.32 (0.72) | 4.35 (0.72) |
| Item 6 | 4.09 (0.87) | 4.11 (0.85) |
| Item 7 | 4.25 (0.63) | 4.29 (0.62) |
| Item 8 | 4.30 (0.59) | 4.31 (0.64) |
| Item 9 | 4.64 (0.64) | 4.67 (0.63) |
| **Esthetics** | 3.33 (0.79) | 3.32 (0.82) |
| Item 10 | 3.73 (0.64) | 3.75 (0.62) |
| Item 11 | 3.26 (0.77) | 3.23 (0.83) |
| Item 12 | 3.00 (0.79) | 3.00 (0.80) |
| **Information quality** | 3.25 (1.08) | 3.25 (1.08) |
| Item 13 | 3.96 (0.39) | 3.96 (0.39) |
| Item 14 | 3.78 (0.70) | 3.75 (0.71) |
| Item 15 | 3.50 (0.75) | 3.49 (0.75) |
| Item 16 | 3.21 (0.80) | 3.22 (0.76) |
| Item 17 | 3.42 (0.98) | 3.43 (0.99) |
| Item 18 | 1.64 (0.90) | 1.66 (0.90) |
| Item 19 | N/A[a] | N/A[a] |
| Overall mean | 3.43 (0.43) | 3.44 (0.43) |

[a]N/A: this item on information quality could not be rated because it was nonapplicable.

## Internal Consistency

The internal consistency of MARS and MARS-F and their subscales is presented in Table 2. The internal consistency of the MARS dimension *engagement* (ω=0.82, 95% CI 0.79-0.87) was good. The internal consistencies of the dimensions *functionality* (ω=0.80, 95% CI 0.74-0.85) and *esthetics* (ω=0.79, 95% CI 0.73-0.88) were acceptable and that for *information quality* (ω=0.64, 95% CI 0.49-0.70) indicated questionable consistency. The internal consistency of the overall MARS score was good (ω=0.87, 95% CI 0.83-0.91).

For MARS-F, the internal consistencies were acceptable for the dimensions *engagement* (ω=0.79, 95% CI 0.72-0.83), *functionality* (ω=0.79, 95% CI 0.73-0.85), *esthetics* (ω=0.78, 95% CI 0.71-0.82), and *information quality* (ω=0.61, 95% CI 0.53-0.65). The internal consistency of the overall MARS score was good (ω=0.86, 95% CI 0.85-0.90).

**Table 2.** Internal consistency per dimension for the original version of the Mobile App Rating Scale (MARS) and the French version of MARS (MARS-F).

| Dimension | Internal consistency, ω (95% CI) | |
|---|---|---|
| | MARS | MARS-F |
| Engagement | 0.82 (0.79-0.87) | 0.79 (0.72-0.83) |
| Functionality | 0.80 (0.74-0.85) | 0.79 (0.73-0.85) |
| Esthetics | 0.79 (0.73-0.88) | 0.78 (0.71-0.82) |
| Information quality | 0.64 (0.49-0.70) | 0.61 (0.53-0.65) |
| Overall mean | 0.87 (0.83-0.91) | 0.86 (0.85-0.90) |

## Validity

The correlation coefficients between the corresponding dimensions of MARS and MARS-F ranged from 0.97 to 0.99. $P$ values were adjusted for multiple testing according to Holmes' method (Table 3). Correlations between the respective items are presented in Multimedia Appendix 2.

**Table 3.** Correlation between the English and French versions of the Mobile App Rating Scale.

| Dimension | Engagement FR | Functionality FR | Esthetics FR | Information quality FR |
|---|---|---|---|---|
| **Engagement ENG** | | | | |
| $r$ | 0.98 | 0.30 | 0.57 | 0.63 |
| $P$ value | <.001 | .02 | <.001 | <.001 |
| **Functionality ENG** | | | | |
| $r$ | 0.24 | 0.98 | 0.49 | 0.28 |
| $P$ value | .03 | <.001 | <.001 | .02 |
| **Esthetics ENG** | | | | |
| $r$ | 0.52 | 0.50 | 0.99 | 0.52 |
| $P$ value | <.001 | <.001 | <.001 | <.001 |
| **Information quality ENG** | | | | |
| $r$ | 0.66 | 0.34 | 0.56 | 0.97 |
| $P$ value | <.001 | .01 | <.001 | <.001 |

## MSA Results

MSA results for both versions of the scale (ie, MARS and MARS-F) are summarized in the Table 4. MSA for MARS revealed strong scalability (H=0.37; SE=0.03). There was no violation of monotonicity because the item step response functions were nondecreasing functions; likewise, there was no violation of nonintersection because the item step response functions do not intersect. The internal consistency of this scale was acceptable (MS=0.88; λ-2=0.88; LCRC=0.89). MSA for MARS-F revealed good scalability (H=0.35; SE=0.03). The internal consistency of this scale was acceptable (MS=0.88; λ-2=0.89; LCRC=0.90). The scalability results of MARS and MARS-F are presented in Figure 2.

**Table 4.** Mokken scale analysis for the original version of the Mobile App Rating Scale (MARS) and the French version of MARS (MARS-F).

| | MARS | MARS-F |
|---|---|---|
| Loevinger H coefficient | 0.37 | 0.35 |
| Standard errors of the scalability coefficients of the item pairs | 0.03 | 0.03 |
| Molenaar-Sijtsma coefficient | 0.88 | 0.88 |
| Lambda-2 | 0.88 | 0.89 |
| Latent class reliability coefficient | 0.89 | 0.90 |

**Figure 2.** Loevinger ($H_k$) coefficients (overall scalability of all items in the scale) for the Mobile App Rating Scale (MARS) and the French version of the Mobile App Rating Scale (MARS-F) depending on various dimensions.

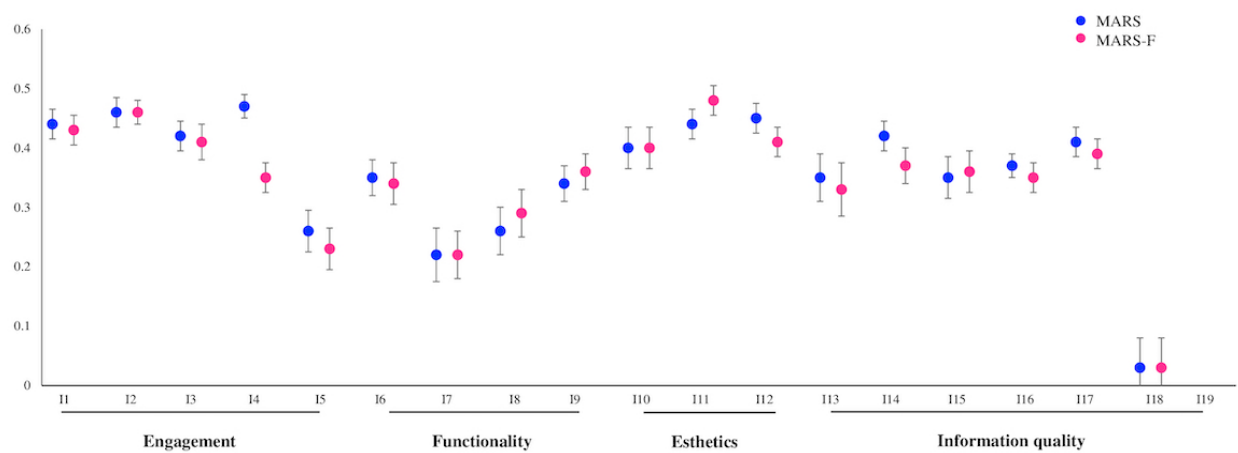

## Discussion

### Principal Results

This study aimed to develop and evaluate MARS-F to enable French health care professionals to assess the quality of mHealth apps. To our knowledge, this is the first cultural adaptation, translation, and validity evaluation of the original MARS in French.

Nutrition-related apps were identified using well-defined and selected search terms in both two app stores (Google Play Store and Apple Store). This was done to avoid methodological challenges such as ranking algorithms or irrelevant results because the indexing of apps is usually determined by a developer who is most interested in promoting the app.

With a view to provide a comparable interpretation of statistical indicators, the methodology was chosen to be similar to previous adaptations of the scale [2,8,14,26]. In addition, 63 apps were included, which is higher than the minimum sample size of 41 apps required to confirm that interrater reliability lies within 0.15 of a sample observation of 0.80, with 87% assurance [26]. We used the same strategy that led to the Italian version of MARS except that the team included apps by searching and screening across three app stores (Google Play, Apple, and Windows Stores). As per the validation of the German version of MARS, each search term was provided separately, as no truncation or use of logic operators (AND, OR, and NOT) was possible in the Google Play and Apple Store. In our study, two raters downloaded and then evaluated 10 apps that were randomly selected for training and piloting purposes as in the initial English version of MARS against 5 apps in the Italian [2], Spanish [13], and German development versions [14].

The interrater reliability of MARS and MARS-F were aligned, with overlapping CI values. The ICCs for MARS-F were also comparable to the Arabic (0.84, 95% CI 0.82-0.85) [15] and German versions of MARS (0.83, 95% CI 0.82-0.85) [14], and they were slightly lower than the Italian version of MARS (0.96, CI 0.90-0.91) [2] (Multimedia Appendix 3).

The internal consistency of the overall MARS score was good and that of MARS-F was acceptable for the dimensions *engagement*, *functionality*, *esthetics*, and *information quality*. The internal consistency of the German version of MARS was good for *engagement* and excellent for *functionality* and *esthetics*. On the other hand, the internal consistency of *information quality* was acceptable. For the Arabic version of MARS, the internal consistency was good for *engagement* and *esthetics*, good for *information quality*, and acceptable for *functionality* [15]. All Cronbach alpha coefficients were judged to be at least acceptable for the Italian version of MARS [2], and these values were high for the Spanish [13] version of MARS.

MSA results for MARS-F revealed a good scalability (H=0.35, SE=0.03), and the use of total MARS-F score was found to be appropriate. Additionally, we obtained a high correspondence between MARS-F and the original MARS [8], which demonstrates proven validity.

The same methodology was used for the validation of the German (apps targeting anxiety), Italian (primary prevention), Spanish (health and fitness apps), and Arabic versions (health and fitness apps). Our results were consistent with the findings of the research teams that developed and validated the Italian, Spanish, German, and Arabic versions of the MARS [2,14,15] (Multimedia Appendix 3).

### Limitations

The first possible limitation could be that the validation of MARS-F is based on the evaluation of nutrition-related apps, whereas MARS is applicable to mHealth apps. The second limitation could be attributed to the fact that MARS-F was elaborated by native speakers living in France. French speakers can have diverse cultures according to their country. Therefore, further adaptation could be required. The third limitation concerns item 19 on information quality. This item could not be rated because raters choose the response option "non applicable," which allows raters to skip an item if the app does not contain any health-related information (eg, nutrition apps in this study). The same item was also excluded from all calculations in the Italian version of MARS because of lack of ratings [2]. This item evaluates the evidence-based literature relating to the nutrition app assessed, and it is worth noting that many apps have not yet been scientifically evaluated.

## Future Perspectives

With 300 billion French-speakers worldwide [27], the translation of MARS could be of special interest. Owing to its wide use in the assessment of mHealth apps in the scientific literature, we chose to translate MARS into French to provide a reliable and understandable tool for health professionals to get an evidence-based sense of the quality and reliability of chosen mHealth apps. Other rating scales such as App Quality Evaluation (AQEL) [28], ENLIGHT [29], and the app evaluation model from the American Psychiatric Association [30] could also represent relevant tools to evaluate mHealth apps for further investigations. All these scales were created for the evaluation of mHealth apps, except AQEL that specifically evaluates nutrition-related apps [28]. Several studies have demonstrated that nutrition is one of the key factors in oral and general health [31]. It would be interesting to translate this scale into French and to evaluate the nutrition-related apps included in our study.

Alongside the assessment process of mHealth apps, the patient's involvement in such processes should also be considered. The user version of the MARS (uMARS) [32] should be translated and evaluated for reliability and validity. Mobile technology represents an innovative opportunity to assist end users in improving their management of their chronic conditions. Such in-the-pocket devices could be adapted to the specific needs of populations. As an example, mHealth apps could be used for young people's transition to adult care services [33], to support active adults [34], or to promote healthy aging [35]. mHealth apps are valuable for the primary and secondary prevention of chronic diseases, especially for controlling individual risk factors and preventing the snowball effect of chronic diseases with aging [31].

## Conclusions

To conclude, MARS-F would be a crucial aid for researchers, health care professionals, public health authorities, and interested third parties, to assess the quality of mHealth apps in French-speaking countries. In addition, French app developers could use this French version as a tool to evaluate and improve the quality of their apps prior to market launch. MARS-F is an important cornerstone to app quality assessment with the purpose to identify reliable and valid apps for the benefit of end users.

### Acknowledgments

We would like to thank Catherine Billard, Adeline Darlington-Bernard, Géraldine Seveste for their contribution in the translation process.

### Conflicts of Interest

None declared.

### Multimedia Appendix 1

French version of the Mobile Application Rating Scale (MARS-F).
[[PDF File (Adobe PDF File), 190 KB](#)-[Multimedia Appendix 1](#)]

### Multimedia Appendix 2

Correlation coefficients between the respective items of the Mobile App Rating Scale (MARS) and the French version of the Mobile App Rating Scale (MARS-F).
[[XLSX File (Microsoft Excel File), 12 KB](#)-[Multimedia Appendix 2](#)]

### Multimedia Appendix 3

Comparison of internal consistencies and the Mokken Scale Analysis between Mobile App Rating Scale in English and other available translations of the scale.
[[PDF File (Adobe PDF File), 57 KB](#)-[Multimedia Appendix 3](#)]

### References

1.  Zhao J, Freeman B, Li M. Can mobile phone apps influence people's health behavior change? an evidence review. J Med Internet Res 2016 Oct 31;18(11):e287 [FREE Full text] [doi: 10.2196/jmir.5692] [Medline: 27806926]
2.  Domnich A, Arata L, Amicizia D, Signori A, Patrick B, Stoyanov S, et al. Development and validation of the Italian version of the Mobile Application Rating Scale and its generalisability to apps targeting primary prevention. BMC Med Inform Decis Mak 2016;16:83 [FREE Full text] [doi: 10.1186/s12911-016-0323-2] [Medline: 27387434]
3.  Rabbi M, Pfammatter A, Zhang M, Spring B, Choudhury T. Automated personalized feedback for physical activity and dietary behavior change with mobile phones: a randomized controlled trial on adults. JMIR Mhealth Uhealth 2015;3(2):e42 [FREE Full text] [doi: 10.2196/mhealth.4160] [Medline: 25977197]

XSL•FO
RenderX

4.  Laing BY, Mangione CM, Tseng C, Leng M, Vaisberg E, Mahida M, et al. Effectiveness of a smartphone application for weight loss compared with usual care in overweight primary care patients: a randomized, controlled trial. Ann Intern Med 2014 Nov 18;161(10 Suppl):S5-12. [doi: [10.7326/M13-3005](10.7326/M13-3005)] [Medline: [25402403](25402403)]
5.  Bort-Roig J, Gilson ND, Puig-Ribera A, Contreras RS, Trost SG. Measuring and influencing physical activity with smartphone technology: a systematic review. Sports Med 2014 May;44(5):671-686. [doi: [10.1007/s40279-014-0142-5](10.1007/s40279-014-0142-5)] [Medline: [24497157](24497157)]
6.  Kampmeijer R, Pavlova M, Tambor M, Golinowska S, Groot W. The use of e-health and m-health tools in health promotion and primary prevention among older adults: a systematic literature review. BMC Health Serv Res 2016 Sep 05;16 Suppl 5:290 [[FREE Full text](FREE Full text)] [doi: [10.1186/s12913-016-1522-3](10.1186/s12913-016-1522-3)] [Medline: [27608677](27608677)]
7.  Ventola CL. Mobile devices and apps for health care professionals: uses and benefits. P T 2014 May;39(5):356-364 [[FREE Full text](FREE Full text)] [Medline: [24883008](24883008)]
8.  Stoyanov SR, Hides L, Kavanagh DJ, Zelenko O, Tjondronegoro D, Mani M. Mobile app rating scale: a new tool for assessing the quality of health mobile apps. JMIR Mhealth Uhealth 2015;3(1):e27 [[FREE Full text](FREE Full text)] [doi: [10.2196/mhealth.3422](10.2196/mhealth.3422)] [Medline: [25760773](25760773)]
9.  Llorens-Vernet P, Miró J. Standards for mobile health-related apps: systematic review and development of a guide. JMIR Mhealth Uhealth 2020 Mar 03;8(3):e13057 [[FREE Full text](FREE Full text)] [doi: [10.2196/13057](10.2196/13057)] [Medline: [32130169](32130169)]
10. Huang H, Bashir M. Users' adoption of mental health apps: examining the impact of information cues. JMIR Mhealth Uhealth 2017 Jun 28;5(6):e83 [[FREE Full text](FREE Full text)] [doi: [10.2196/mhealth.6827](10.2196/mhealth.6827)] [Medline: [28659256](28659256)]
11. Singh K, Drouin K, Newmark LP, Lee J, Faxvaag A, Rozenblum R, et al. Many mobile health apps target high-need, high-cost populations, but gaps remain. Health Aff (Millwood) 2016 Dec 01;35(12):2310-2318. [doi: [10.1377/hlthaff.2016.0578](10.1377/hlthaff.2016.0578)] [Medline: [27920321](27920321)]
12. Azad-Khaneghah P, Neubauer N, Miguel Cruz A, Liu L. Mobile health app usability and quality rating scales: a systematic review. Disabil Rehabil Assist Technol 2020 Jan 08:1-10. [doi: [10.1080/17483107.2019.1701103](10.1080/17483107.2019.1701103)] [Medline: [31910687](31910687)]
13. Martin Payo R, Fernandez Álvarez MM, Blanco Díaz M, Cuesta Izquierdo M, Stoyanov SR, Llaneza Suárez E. Spanish adaptation and validation of the Mobile Application Rating Scale questionnaire. Int J Med Inform 2019 Sep;129:95-99. [doi: [10.1016/j.ijmedinf.2019.06.005](10.1016/j.ijmedinf.2019.06.005)] [Medline: [31445295](31445295)]
14. Messner E, Terhorst Y, Barke A, Baumeister H, Stoyanov S, Hides L, et al. The German version of the Mobile App Rating Scale (MARS-G): development and validation study. JMIR Mhealth Uhealth 2020 Mar 27;8(3):e14479 [[FREE Full text](FREE Full text)] [doi: [10.2196/14479](10.2196/14479)] [Medline: [32217504](32217504)]
15. Bardus M, Awada N, Ghandour LA, Fares E, Gherbal T, Al-Zanati T, et al. The Arabic version of the Mobile App Rating Scale: development and validation study. JMIR Mhealth Uhealth 2020 Mar 03;8(3):e16956 [[FREE Full text](FREE Full text)] [doi: [10.2196/16956](10.2196/16956)] [Medline: [32130183](32130183)]
16. Herdman M, Fox-Rushby J, Badia X. A model of equivalence in the cultural adaptation of HRQoL instruments: the universalist approach. Qual Life Res 1998 May;7(4):323-335. [doi: [10.1023/a:1024985930536](10.1023/a:1024985930536)] [Medline: [9610216](9610216)]
17. Portney L, Watkins M. Foundations of Clinical Research: Applications to Practice. United States: F.A. Davis Company; 2009.
18. McNeish D. Thanks coefficient alpha, we'll take it from here. Psychol Methods 2018 Sep;23(3):412-433. [doi: [10.1037/met0000144](10.1037/met0000144)] [Medline: [28557467](28557467)]
19. Zhang Z, Yuan K. Robust coefficients alpha and omega and confidence intervals with outlying observations and missing data: methods and software. Educ Psychol Meas 2016 Jun;76(3):387-411 [[FREE Full text](FREE Full text)] [doi: [10.1177/0013164415594658](10.1177/0013164415594658)] [Medline: [29795870](29795870)]
20. Holm S. A simple sequentially rejective multiple test procedure. Scandinavian Journal of Statistics 1979;6(2):65-70 [[FREE Full text](FREE Full text)]
21. Ark LAVD. Mokken Scale Analysis in R. J Stat Soft 2007;20(11):1-19. [doi: [10.18637/jss.v020.i11](10.18637/jss.v020.i11)]
22. Sijtsma K, van der Ark LA. A tutorial on how to do a Mokken scale analysis on your test and questionnaire data. Br J Math Stat Psychol 2017 Feb;70(1):137-158. [doi: [10.1111/bmsp.12078](10.1111/bmsp.12078)] [Medline: [27958642](27958642)]
23. Ark LAVD. New developments in Mokken Scale Analysis in R. J Stat Soft 2012;48(5):1-27 [[FREE Full text](FREE Full text)] [doi: [10.18637/jss.v048.i05](10.18637/jss.v048.i05)]
24. Molenaar I, Sijtsma K. Mokken's approach to reliability estimation extended to multicategory items. Kwantitatieve Methoden: Nieuwsbrief voor Toegepaste Statistiek en Operationele Research? 1988;9(28):115-126.
25. van der Ark LA, van der Palm DW, Sijtsma K. A latent class approach to estimating test-score reliability. Applied Psychological Measurement 2011 Mar 09;35(5):380-392. [doi: [10.1177/0146621610392911](10.1177/0146621610392911)]
26. Zou GY. Sample size formulas for estimating intraclass correlation coefficients with precision and assurance. Stat Med 2012 Dec 20;31(29):3972-3981. [doi: [10.1002/sim.5466](10.1002/sim.5466)] [Medline: [22764084](22764084)]
27. Qui parle français dans le monde – Organisation internationale de la Francophonie. Webpage in French. Langue française et diversité linguistique. URL: [http://observatoire.francophonie.org/qui-parle-francais-dans-le-monde/](http://observatoire.francophonie.org/qui-parle-francais-dans-le-monde/) [accessed 2021-05-03]
28. DiFilippo KN, Huang W, Chapman-Novakofski KM. A new tool for nutrition App Quality Evaluation (AQEL): development, validation, and reliability testing. JMIR Mhealth Uhealth 2017 Oct 27;5(10):e163 [[FREE Full text](FREE Full text)] [doi: [10.2196/mhealth.7441](10.2196/mhealth.7441)] [Medline: [29079554](29079554)]

29. Baumel A, Faber K, Mathur N, Kane JM, Muench F. Enlight: a comprehensive quality and therapeutic potential evaluation tool for mobile and web-based eHealth interventions. J Med Internet Res 2017 Mar 21;19(3):e82 [FREE Full text] [doi: 10.2196/jmir.7270] [Medline: 28325712]
30. The App Evaluation Model. American Psychiatric Association. URL: https://www.psychiatry.org/psychiatrists/practice/mental-health-apps/the-app-evaluation-model [accessed 2021-05-03]
31. Martinon P, Fraticelli L, Giboreau A, Dussart C, Bourgeois D, Carrouel F. Nutrition as a key modifiable factor for periodontitis and main chronic diseases. J Clin Med 2021 Jan 07;10(2):197 [FREE Full text] [doi: 10.3390/jcm10020197] [Medline: 33430519]
32. Stoyanov SR, Hides L, Kavanagh DJ, Wilson H. Development and validation of the user version of the Mobile Application Rating Scale (uMARS). JMIR Mhealth Uhealth 2016 Jun 10;4(2):e72 [FREE Full text] [doi: 10.2196/mhealth.5849] [Medline: 27287964]
33. Virella Pérez YI, Medlow S, Ho J, Steinbeck K. Mobile and web-based apps that support self-management and transition in young people with chronic illness: systematic review. J Med Internet Res 2019 Nov 20;21(11):e13579 [FREE Full text] [doi: 10.2196/13579] [Medline: 31746773]
34. Mascarenhas MN, Chan JM, Vittinghoff E, Van Blarigan EL, Hecht F. Increasing physical activity in mothers using video exercise groups and exercise mobile apps: randomized controlled trial. J Med Internet Res 2018 May 18;20(5):e179 [FREE Full text] [doi: 10.2196/jmir.9310] [Medline: 29776899]
35. Chen YR, Schulz PJ. The effect of information communication technology interventions on reducing social isolation in the elderly: a systematic review. J Med Internet Res 2016 Jan 28;18(1):e18 [FREE Full text] [doi: 10.2196/jmir.4596] [Medline: 26822073]

## Abbreviations

**AQEL:** app quality evaluation
**ICC:** intraclass correlation coefficients
**LCRC:** latent class reliability coefficient
**MARS:** Mobile Application Rating Scale
**MARS-F:** Mobile Application Rating Scale–French
**MSA:** Mokken Scale Analysis
**MS:** Molenaar-Sijtsma
**uMARS:** user version of the Mobile Application Rating Scale